\documentclass[aps,prc,twocolumn,superscriptaddress,floatfix]{revtex4} 

\usepackage{graphicx} 
\usepackage{bm} 
\usepackage[dvipsnames]{xcolor}
\usepackage{scrextend} 
\usepackage{longtable} 
\usepackage[tight-spacing=true]{siunitx} 
\makeatletter
\newcommand*{\centerfloat}{%
  \parindent \z@
  \leftskip \z@ \@plus 1fil \@minus \textwidth
  \rightskip\leftskip
  \parfillskip \z@skip}
\makeatother

\begin{document}

\title{Detailed low-spin spectroscopy of $^{65}$Ni via neutron capture reaction}

\author{C.~Porzio\footnote{\label{1}Corresponding author: carlotta.porzio@mi.infn.it}} 
\affiliation{Dipartimento di Fisica, Universit\`a degli Studi di Milano, via Celoria 16, 1-20133, Milano, Italy} 
\affiliation{INFN sezione di Milano, via Celoria 16, 1-20133, Milano, Italy} 
\affiliation{Institut Laue-Langevin, 71 Avenue des Martyrs, F-38042 Grenoble, France}
\author{C.~Michelagnoli}
\affiliation{Institut Laue-Langevin, 71 Avenue des Martyrs, F-38042 Grenoble, France}
\author{N.~Cieplicka-Ory\'{n}czak}
\affiliation{Institute of Nuclear Physics, Polish Academy of Sciences, PL-31342 Krak\'{o}w, Poland}
\author{M.~Sferrazza}
\affiliation{D\'{e}partement de Physique, Facult\'{e} des Sciences, Université Libre de Bruxelles, Boulevard du Triomphe, Bruxelles 1050, Belgium}
\author{S.~Leoni\footnote{\label{2}Corresponding author: silvia.leoni@mi.infn.it}} 
\affiliation{Dipartimento di Fisica, Universit\`a degli Studi di Milano, via Celoria 16, 1-20133, Milano, Italy} 
\affiliation{INFN sezione di Milano, via Celoria 16, 1-20133, Milano, Italy} 
\author{B.~Fornal}
\affiliation{Institute of Nuclear Physics, Polish Academy of Sciences, PL-31342 Krak\'{o}w, Poland}
\author{Y.~Tsunoda}
\affiliation{Center for Nuclear Study, The University of Tokyo, 7-3-1 Hongo, Bunkyo-ku, Tokyo 113-0033, Japan}
\author{T.~Otsuka}
\affiliation{Center for Nuclear Study, The University of Tokyo, 7-3-1 Hongo, Bunkyo-ku, Tokyo 113-0033, Japan}
\affiliation{RIKEN Nishina Center, Wako, Saitama 351-0198, Japan}
\affiliation{National Superconducting Cyclotron Laboratory, Michigan State University, East Lansing, Michigan 48824, USA}
\affiliation{Instituut voor Kern- en Stralingsfysica, Katholieke Universiteit Leuven, Leuven B-3001, Belgium}
\author{S.~Bottoni}
\affiliation{Dipartimento di Fisica, Universit\`a degli Studi di Milano, via Celoria 16, 1-20133, Milano, Italy} 
\affiliation{INFN sezione di Milano, via Celoria 16, 1-20133, Milano, Italy} 
\author{C.~Costache}
\affiliation{Horia Hulubei National Institute of Physics and Nuclear Engineering IFIN-HH, Bucharest 077125, Romania}
\author{F.~C.~L.~Crespi}
\affiliation{Dipartimento di Fisica, Universit\`a degli Studi di Milano, via Celoria 16, 1-20133, Milano, Italy} 
\affiliation{INFN sezione di Milano, via Celoria 16, 1-20133, Milano, Italy} 
\author{ \L{}.~W.~Iskra}
\affiliation{INFN sezione di Milano, via Celoria 16, 1-20133, Milano, Italy} 
\author{M.~Jentschel}
\affiliation{Institut Laue-Langevin, 71 Avenue des Martyrs, F-38042 Grenoble, France}
\author{F.~Kandzia}
\affiliation{Institut Laue-Langevin, 71 Avenue des Martyrs, F-38042 Grenoble, France}
\author{Y.-H.~Kim}
\affiliation{Institut Laue-Langevin, 71 Avenue des Martyrs, F-38042 Grenoble, France}
\author{U.~K\"{o}ster}
\affiliation{Institut Laue-Langevin, 71 Avenue des Martyrs, F-38042 Grenoble, France}
\author{N.~M\u{a}rginean}
\affiliation{Horia Hulubei National Institute of Physics and Nuclear Engineering IFIN-HH, Bucharest 077125, Romania}
\author{C.~Mihai}
\affiliation{Horia Hulubei National Institute of Physics and Nuclear Engineering IFIN-HH, Bucharest 077125, Romania}
\author{P.~Mutti}
\affiliation{Institut Laue-Langevin, 71 Avenue des Martyrs, F-38042 Grenoble, France}
\author{A.~Turturic\u{a}}
\affiliation{Horia Hulubei National Institute of Physics and Nuclear Engineering IFIN-HH, Bucharest 077125, Romania}      

\date{\today}

\begin{abstract}
An extended investigation of the low-spin structure of the $^{65}$Ni nucleus was performed at the Institut Laue-Langevin, Grenoble,  via the neutron capture reaction $^{64}$Ni(n,$\gamma$)$^{65}$Ni, using the FIPPS  HPGe array.
The level scheme of $^{65}$Ni was significantly expanded, with 2 new levels and 87 newly found transitions. Angular correlation analyses were also performed, allowing us to assign spins and parities for a number of states, and to determine multipolarity mixing ratios for selected $\gamma$~transitions. 
The low-energy part of the experimental level scheme (up to about 1.4~MeV) was compared with Monte Carlo Shell Model calculations, which predict spherical shapes for all states, apart from the 9/2$^+$ and the second excited 1/2$^-$ states of oblate deformation.
\end{abstract}


\maketitle


\section{\label{sec:intro}Introduction}

Neutron capture reactions, later referred as (n,$\gamma$), populate the nucleus at the neutron separation energy and rather low spin (i.e., the spin of the target nucleus coupled to the 1/2 spin of the neutron). Such system deexcites via $\gamma$-ray cascades towards the ground state. Using $\gamma$-coincidence techniques, a detailed level scheme of low-spin states, which is complementary to the ones reconstructed via fusion, deep-inelastic or transfer reactions, can be obtained.
In neutron-rich Ni isotopes, detailed studies of the low-spin structures are particularly relevant, since they may shed light on the shape coexistence phenomenon. This phenomenon
has been predicted, at low spin, in these nuclei by various theoretical calculations~\cite{Girod,Bonche,Moeller2009,Moeller2012,Tsunoda,Otsuka} and also clearly observed in the even-even Ni isotopes with mass $A$ = 64, 66, 68 and 70~\cite{Crider,Gade2016,Leoni,Olaizola,Stryjczyk,Marginean2020}. 

In this context, the odd system $^{65}$Ni is of special interest  since it sits in between the even-even systems $^{66}$Ni, where a shape-isomer-like structure (an extreme case of shape coexistence) has been recently observed~\cite{Leoni}, and $^{64}$Ni, which also shows coexistence of spherical, oblate and prolate shapes~\cite{Marginean2020}. A shape coexistence scenario is, therefore, expected also in $^{65}$Ni, although with a more complex pattern due to its even-odd nature.

Our knowledge on the structure of $^{65}$Ni comes mainly from (n,$\gamma$) experiments performed about 40 years ago~\cite{Arnell,Cochavi,Ishaq,Vennink}, deep-inelastic scattering experiments from the 90's~\cite{Pawlat}, (d,p) reaction studies~\cite{Anfinsen,Schiffer2013,Georgiev2006} and $\beta$-decay studies~\cite{Pauwels2009}. The aim of this work is to perform a complete low-spin spectroscopic study of $^{65}$Ni, by using a (n,$\gamma$) reaction, in order to significantly expand our understanding of its low-spin structure and search for signature of shape coexistence.

The experiment was performed at the reactor of Institut Laue-Langevin (ILL Grenoble, France), using the array FIPPS (FIssion Product Prompt gamma-ray Spectrometer). The level scheme of $^{65}$Ni was significantly expanded, with 2 new levels and 87 newly found transitions. In addition, angular correlation studies allowed us to determine the multipolarity of several transitions and firmly assign spin and parity to a number of states.

The low energy part of the experimental level scheme, up to around 1.4 MeV, was compared with the results from Monte Carlo Shell Model (MCSM) calculations, which provide a description of the structure of the states in terms of wave function compositions and associated intrinsic shapes~\cite{Tsunoda,Otsuka,Togashi}. A dominance of spherical configurations is observed in this energy range, with one possible excitation of oblate nature.
The paper is organized as follows: Section~\ref{sec:exp} contains the experimental details and results, including the construction of the level scheme (Sec.~\ref{sec:levsch}), the $\gamma$-transition intensity determination (Sec.~\ref{sec:intensities}) and the $\gamma$-angular correlation analysis (Sec.~\ref{sec:angcorr}). In Section~\ref{sec:theor}, a comparison between the experimental level scheme and MCSM shell-model predictions is presented. Section~\ref{sec:conclus} contains the conclusions.

\section{\label{sec:exp}Experimental method and results}
\begin{figure*}
\centering
\includegraphics[width=\textwidth]{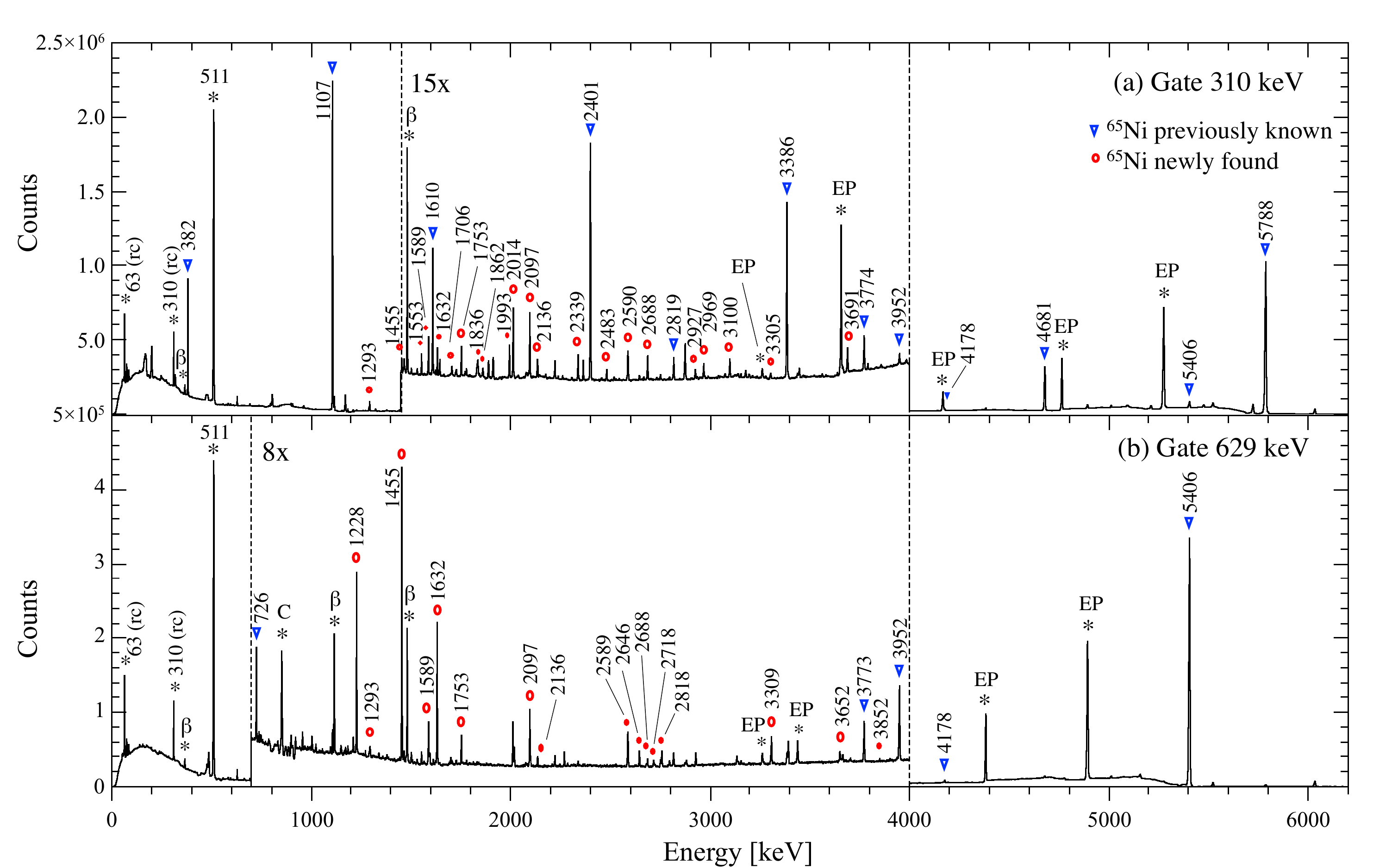}
\caption{Coincidence energy spectra gated on the 310 and 629~keV transitions, respectively, are shown in panels (a) and (b). Newly observed transitions are indicated by red circles, while known ones by blue triangles. Stars indicate escape peaks (EP), radiation from $\beta^-$ decay of $^{65}$Ni ($\beta$), random coincidences with $^{65}$Ni itself (rc) and spurious peaks from Compton events~(C), caused by the absence of Compton shields. }
\label{fig:specLowGate} 
\end{figure*}

The experiment was performed at the Institut Laue-Langevin (ILL) in Grenoble, France, using the FIssion Product Prompt gamma-ray Spectrometer (FIPPS) instrument. 
The FIPPS array consisted of 8~HPGe clovers, each comprised of four n-type HPGe crystals with a diameter of 50~mm (before shaping) and a length of 80~mm, which are front tapered~\citep{Michelagnoli}.
The detectors were arranged symmetrically  around the target position perpendicularly to the beam direction.

The $^{65}$Ni nucleus was populated via a thermal neutron capture reaction on a highly isotopically enriched $^{64}$Ni target. The neutron beam was delivered from the ILL reactor, collimated to a diameter of about 1.5~cm at target position and with a flux of 7.7$\times$10$^7$~neutrons~cm$^{-2}$s$^{-1}$. 
The $^{64}$Ni target (86.7~mg, enriched at 99.6\%) was in powder form and was placed in a double-layered teflon bag at a distance of 9~cm from the front face of the HPGe clovers.
The target holder consisted of a cylinder of about 50~mm diameter and 50~cm length, made of lithium fluoride plastic (enriched in $^6$Li). This was used as a shielding for scattered neutrons, thus contributing to the suppression of beam-induced background.

The signals were treated with digital electronics and the data collected in trigger-less mode. The events were built using a 300~ns time window, validated by channel multiplicity 2. A total number of $1.3\times10^{10}$ $\gamma\gamma$ coincidences were collected and sorted into matrices.

\subsection{\label{sec:levsch}Construction of the level scheme}

To build the presented level scheme, the $\gamma\gamma$ coincidence technique was applied. A $\gamma\gamma$ matrix was constructed, considering a coincidence window of 300~ns. The energy calibration was performed using the neutron capture reaction $^{27}$Al(n,$\gamma$)$^{28}$Al, providing reference prompt $\gamma$-rays in an energy range from tens of keV to $\approx$7~MeV. 

The calibrated data were sorted into $\gamma\gamma$ matrices and coincidence gates on known transitions were used to obtain the new spectroscopic information. 76 new transitions were observed  between known states. In Figs.~\ref{fig:specLowGate}(a),(b) the spectra in coincidence with the 310-keV ($3/2^-\rightarrow5/2^-$) and 629-keV ($3/2^-\rightarrow1/2^-$) $\gamma$ rays are shown. The newly found transitions are labelled in red, while the previously known ones are in blue.

\begin{figure}
\centering
\includegraphics[width=\linewidth]{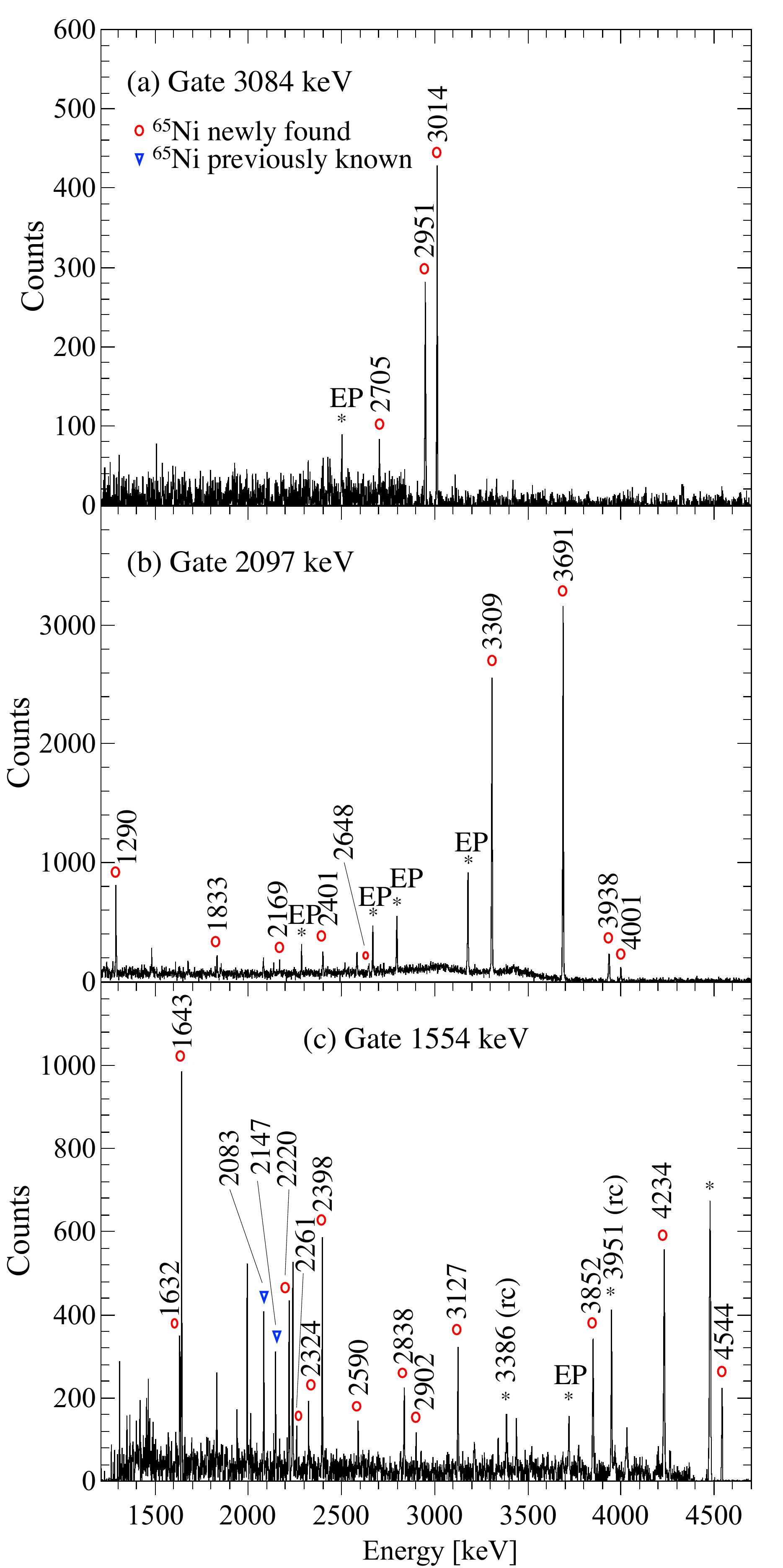}
\caption{Examples of coincidence gated spectra for~$^{65}$Ni, obtained by setting gates on the primary $\gamma$ rays of (a) 3084~keV, (b) 2097~keV and (c) 1554~keV (see level scheme in Figs.~\ref{fig:LevSch1}-\ref{fig:LevSch3}). Escape peaks and peaks from random coincidences are marked with a star. } 
\label{fig:specPrimaryGate} 
\end{figure}

Two new excited levels were identified at 3014.43(14)~keV and 4001.49(18)~keV. 
Spectra gated on the 3084- and 2097-keV transitions populating these levels are presented in Figs.~\ref{fig:specPrimaryGate}(a),(b). They show peaks corresponding to different branches of their decay. In total, 11 new transitions populating or depopulating the two new levels were found. Fig.~\ref{fig:specPrimaryGate}(c) shows also the spectrum gated on the 1554-keV transition which populates the 4544-keV level.

\begin{figure*} [tbhp]
\centering
\includegraphics[width=\textwidth]{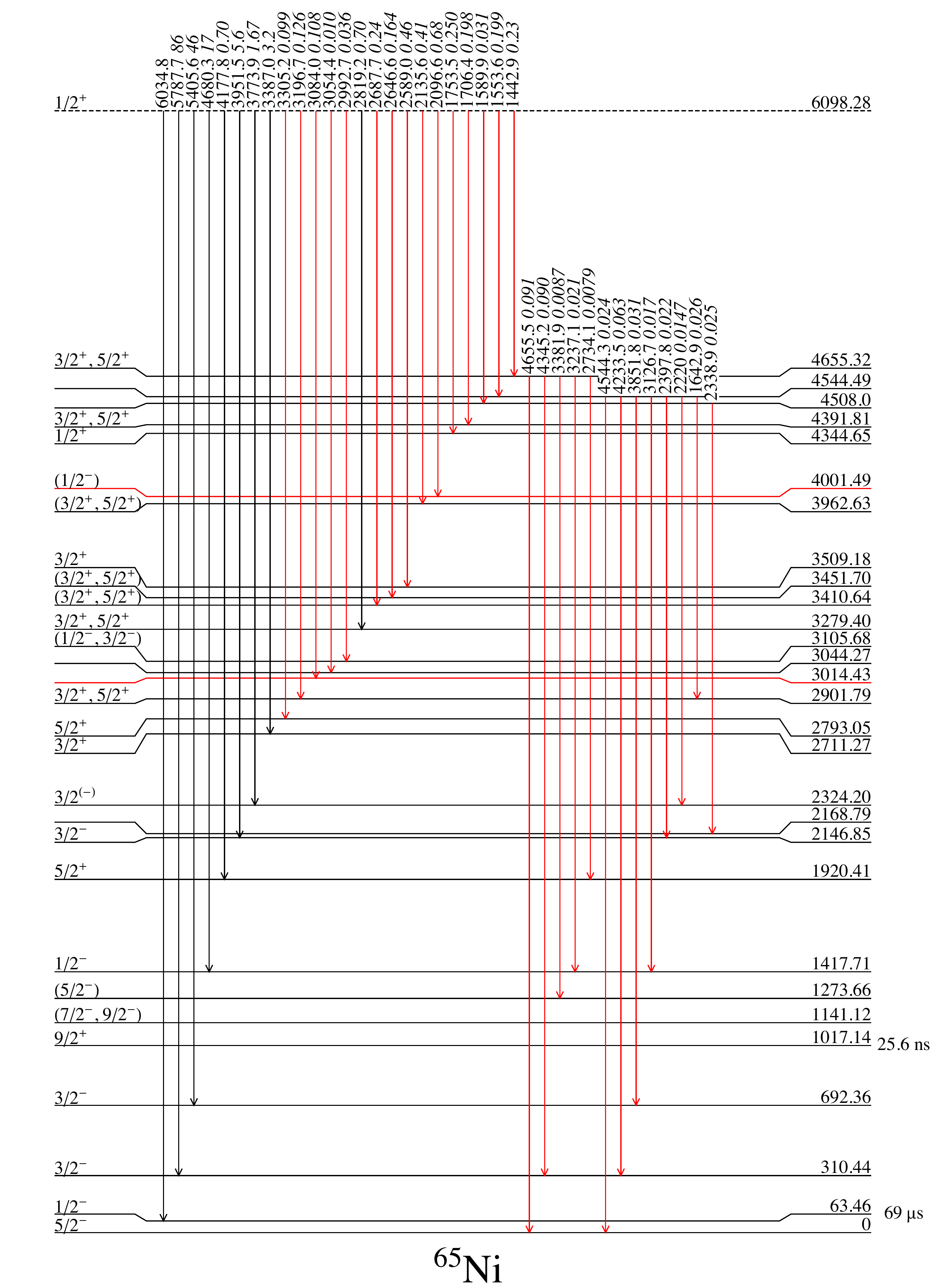}
\caption{First part of the proposed level scheme for $^{65}$Ni. Newly found levels and $\gamma$ transitions are marked in red. Intensities are reported in italic, above transition energies. }
\label{fig:LevSch1} 
\end{figure*}

\begin{figure*} [tbhp]
\centering
\includegraphics[width=\textwidth]{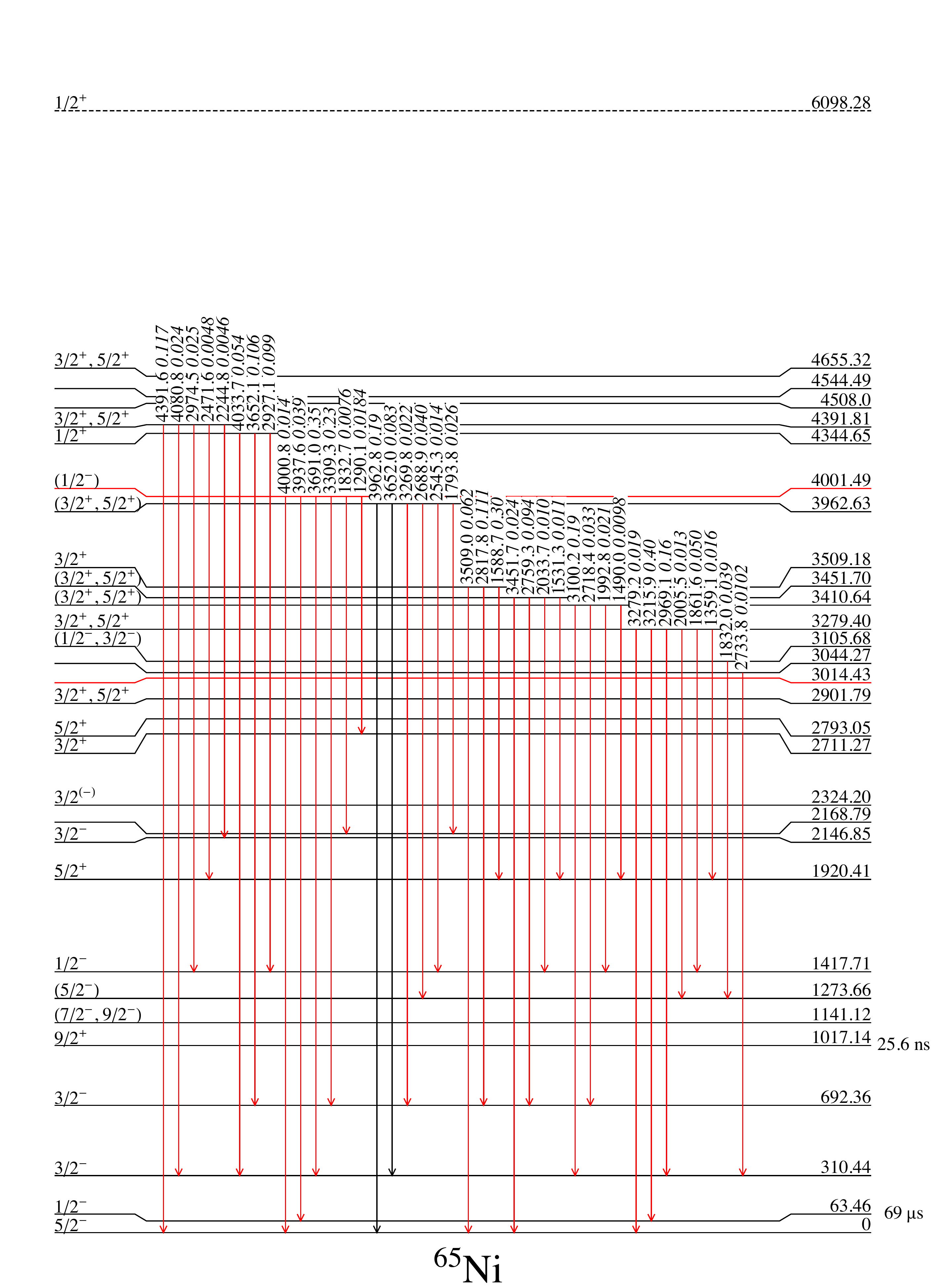}
\caption{Second part of the proposed level scheme for $^{65}$Ni. Newly found levels and $\gamma$ transitions are marked in red. Intensities are reported in italic, above transition energies.}
\label{fig:LevSch2} 
\end{figure*}

\begin{figure*} [tbhp]
\centering
\includegraphics[width=\textwidth]{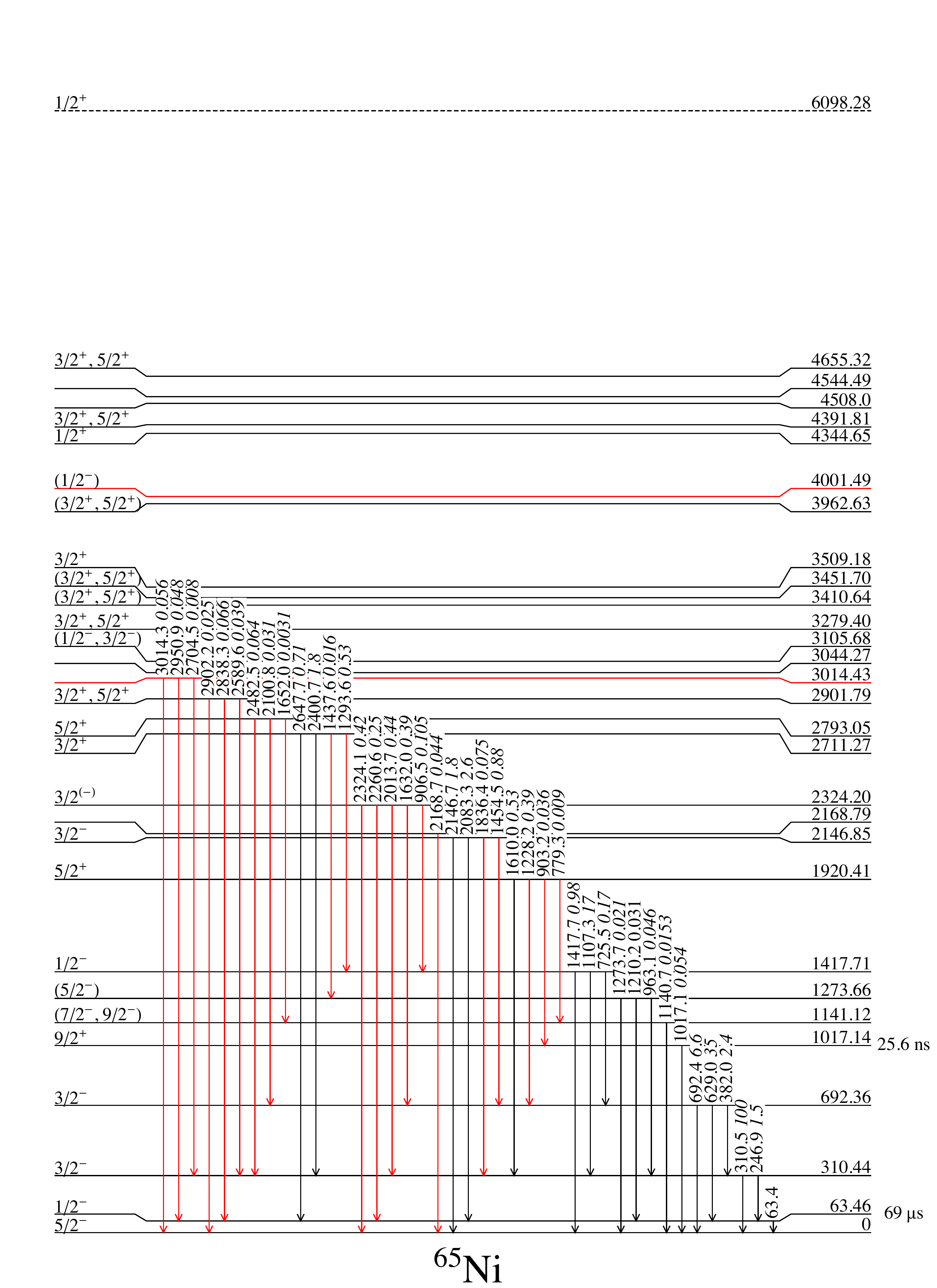}
\caption{Third part of the proposed level scheme for $^{65}$Ni. Newly found levels and $\gamma$ transitions are marked in red. Intensities are reported in italic, above transition energies.}
\label{fig:LevSch3} 
\end{figure*}

The $^{65}$Ni level scheme established in this work is shown in Figs.~\ref{fig:LevSch1}--\ref{fig:LevSch3}. Levels and transitions marked in black were previously known, while the ones in red were identified in the present investigation. Table~\ref{tab:GammaEne} summarizes the information about energy levels and $\gamma$ transitions of $^{65}$Ni, discussed in this work. 


\subsection{\label{sec:intensities}Intensities}

In order to determine the relative $\gamma$-ray intensities following the (n,$\gamma$) reaction,  a detection efficiency curve~$\epsilon_\gamma$ of the FIPPS array was carefully evaluated. Transitions from a $^{152}$Eu calibration source, together with transitions of known intensities from the neutron capture reactions $^{48}$Ti(n,$\gamma$)$^{49}$Ti and $^{27}$Al(n,$\gamma$)$^{28}$Al, were used for this purpose. The efficiency curve was fit with the following phenomenological function:
\begin{equation}
\epsilon_\gamma(E) = \exp\Big( a_1 + a_2E + a_3 \frac{\ln E/E_0}{E} + \frac{a_4}{E} + \frac{a_5}{E^2} + \frac{a_6}{E^3} \Big)
\label{eq:effic}
\end{equation}
where $E=E_\gamma/20$ and $E_0=1$~keV. The parameters resulting from the fit are $a_1=-3.84(8)$, $a_2=-0.0046(4)$~keV$^{-1}$, $a_3=16(2)$~keV, $a_4=-23(4)$~keV, $a_5=43(9)$~keV$^2$, $a_6=-28(5)$~keV$^3$. The uncertainty on the efficiency curve was calculated imposing the condition that about 68\% of the data points should be within one sigma from the fit, considering separately the ranges below and above 1~MeV. 

Relative intensities were obtained from $\gamma\gamma$ coincidence spectra, by using different gating conditions. For each $\gamma$ ray depopulating the neutron capture state (primary $\gamma$ ray), gates were set on all the observed transitions from the level which is fed by the primary transition. It was assumed that they represent the full decaying intensity, owing to the high sensitivity of the FIPPS array.  The relative intensity $I_{\gamma}$ of the considered primary $\gamma$ ray was obtained using eq.~\ref{eq:intensity},
\begin{equation}
I_\gamma = N \sum_i \frac{A_{i}}{\epsilon_\gamma \epsilon_{\gamma_i}} 
\label{eq:intensity}
\end{equation}
where $A_{i}$ is the primary transition peak area when gating on the  transition $i$, $\epsilon_\gamma$ and $\epsilon_{\gamma_i}$ are the singles photopeak efficiencies of the primary and gating $\gamma$ rays. For secondary $\gamma$ rays, gates were set on all the transitions feeding the level of interest and eq.~\ref{eq:intensity} was used. In this case $\epsilon_\gamma$ and $\epsilon_{\gamma_i}$ are the singles photopeak efficiencies of the considered secondary transition and the feeding $\gamma$ rays, respectively. N~is an overall normalization factor calculated by summing the intensities of $\gamma$-$\gamma$ pairs containing the 310-keV ($3/2^-\rightarrow5/2^-$) transition (i.e.,the most intense one observed in the $\gamma\gamma$~matrix) and transitions feeding the 310-keV level.
 
Results of the intensity measurements are listed in Table~\ref{tab:GammaEne}.

\LTcapwidth=\textwidth 
\renewcommand\baselinestretch{1.2}\selectfont  

\clearpage
\begin{longtable*}{@{\extracolsep{\fill}}lcScSS} 
\caption{\label{tab:GammaEne}Information about energy levels and $\gamma$ transitions obtained in the present work. In the first four columns, energies and spin-parities of the initial and final states are given. The excitation energies were obtained through a least-squares fit of all the $\gamma$-ray transition energies established in this work; these are listed in column 5. Spin-parity values are taken from the literature~\citep{ENSDFurl}, except for the ones determined in this work and marked with an asterisk (see Section~\ref{sec:angcorr}). Newly identified levels are labeled as $new$.  The last column shows $\gamma$-ray intensities, relatively to the most intense transition of 310.45~keV in the $\gamma\gamma$~matrix.}\\
\hline
\hline
\textbf{$\boldsymbol{E_{i}}$~[keV]} & \textbf{$\boldsymbol{J^\pi_{i}}$} & {\textbf{$\boldsymbol{E_{f}}$~\textbf{[keV]}}} & \textbf{$\boldsymbol{J^\pi_{f}}$} & {\textbf{$\boldsymbol{E_\gamma}$~\textbf{[keV]}}} & {\textbf{$\boldsymbol{I_\gamma}$}}\\
\hline
\endfirsthead
\multicolumn{6}{c}%
{\tablename\ \thetable\ -- \textit{Continued from previous page}} \\
\hline
\textbf{$\boldsymbol{E_{i}}$~[keV]} & \textbf{$\boldsymbol{J^\pi_{i}}$} & {\textbf{$\boldsymbol{E_{f}}$~\textbf{[keV]}}} & \textbf{$\boldsymbol{J^\pi_{f}}$} & {\textbf{{$\boldsymbol{E_\gamma}$~\textbf{[keV]}}}} & {\textbf{$\boldsymbol{I_\gamma}$}}\\
\hline
\endhead
\hline \multicolumn{6}{r}{\textit{Continued on next page}} \\
\endfoot
\hline
\endlastfoot
63.46(6) & 1/2$^-$ & 0 & 5/2$^-$ & 63.4(1) &  \\
310.44(5)& 3/2$^-$ & 63.46(6) & 1/2$^-$  & 246.9(1) & 1.5(2) \\
		 &  & 0 & 5/2$^-$  & 310.5(1) & 100 \\
692.36(5)& 3/2$^-$ & 310.44(5) & 3/2$^-$ & 382.0(1) & 2.4(3) \\
		 &  & 63.46(6) & 1/2$^-$ & 629.0(1) & 35(5) \\
		 &  & 0 & 5/2$^-$ & 692.4(1) & 6.6(10) \\
1017.14(8) & 9/2$^+$ & 0 & 5/2$^-$ & 1017.1(1) & 0.054(7)\\
1141.12(12)& (7/2$^-$,9/2$^-$) & 0 & 5/2$^-$ & 1140.7(8) & 0.0153(13)\\
1273.66(15)& (5/2$^-$) & 310.44(5) & 3/2$^-$ & 963.1(6) & 0.046(4)\\
		 &  & 63.46(6) & 1/2$^-$ & 1210.2(3) & 0.031(2)\\
		 &  & 0 & 5/2$^-$ & 1273.7(3) & 0.021(2)\\
1417.71(7) & 1/2$^-$ & 692.36(5)& 3/2$^-$ & 725.5(3) & 0.17(2) \\
		 &  & 310.44(5) & 3/2$^-$& 1107.3(1) & 17(2) \\
		 &  & 0 & 5/2$^-$ & 1417.7(2) & 0.98(11) \\
1920.41(8)& 5/2$^+$ & 1141.12(12) & (7/2$^-$,9/2$^-$) & 779.3(1) & 0.009(2) \\
		 &  &  1017.14(8) & 9/2$^+$ & 903.2(1) & 0.036(4) \\
		 &  &  692.36(5) & 3/2$^-$ & 1228.2(2) & 0.39(3) \\
		 &  &  310.44(5) & 3/2$^-$ & 1610.0(1) & 0.53(4) \\
2146.85(15)& 3/2$^-$ & 692.36(5) & 3/2$^-$ & 1454.5(2) & 0.88(9) \\
		 &  & 310.44(5) & 3/2$^-$ & 1836.4(3) & 0.075(9) \\
		 &  & 63.46(6) & 1/2$^-$ & 2083.3(5) & 2.6(3)\\		 
		 &  & 0 & 5/2$^-$ & 2146.7(8) &  1.8(2)\\		 
2168.79(18)&  & 0  & 5/2$^-$ & 2168.7(2) & 0.044(2)\\
2324.20(6)& 3/2$^{(-)}$*  & 1417.71(7) & 1/2$^-$ & 906.5(2) & 0.105(15) \\
		 &  & 692.36(5) & 3/2$^-$ & 1632.0(1) & 0.39(4) \\
		 &  & 310.44(5) & 3/2$^-$ & 2013.7(1) & 0.44(4) \\
		 &  & 63.46(6) & 1/2$^-$  & 2260.6(1) & 0.25(3) \\
		 &  & 0 & 5/2$^-$ & 2324.1(1) & 0.42(4) \\
2711.27(8)& 3/2$^+$ & 1417.71(7) & 1/2$^-$ & 1293.6(1) & 0.53(5) \\
		 &  & 1273.66(15) & (5/2$^-$) & 1437.6(5) & 0.016(2) \\
		 &  & 310.44(5) & 3/2$^-$ & 2400.7(1) & 1.8(2) \\
		 &  & 63.46(6) & 1/2$^-$ & 2647.7(1) & 0.71(7) \\
2793.05(10)& 5/2$^+$ & 1141.12(12)& (7/2$^-$,9/2$^-$) & 1652.0(2) & 0.0031(11) \\
		 &  & 692.36(5) & 3/2$^-$ & 2100.8(4) & 0.031(4)\\
		 &  & 310.44(5) & 3/2$^-$ & 2482.5(1) & 0.064(6)\\
2901.79(11)& 3/2$^+$,5/2$^+$ & 310.44(5) & 3/2$^-$ & 2589.6(8) & 0.039(5) \\
		 &  & 63.46(6) & 1/2$^-$ & 2838.3(1) & 0.066(6) \\
		 &  & 0 & 5/2$^-$ & 2902.2(3) & 0.025(3) \\
3014.43(14) $new$ & & 310.44(5) & 3/2$^-$ & 2704.5(6) & 0.008(2) \\
		 &  & 63.46(6) & 1/2$^-$ & 2950.9(2) & 0.048(5) \\
		 &  & 0 & 5/2$^-$ & 3014.3(2) & 0.056(6) \\
3044.27(20)&  & 310.44(5) & 3/2$^-$ & 2733.8(2) & 0.0102(15) \\
3105.68(24)& (1/2$^-$,3/2$^-$) & 1273.66(15) & (5/2$^-$) & 1832.0(2) & 0.039(4)\\
3279.40(10)& 3/2$^+$,5/2$^+$  &1920.41(8)& 5/2$^+$ & 1359.1(4) &  0.016(2)\\
		 &  & 1417.71(7) & 1/2$^-$ & 1861.6(2) & 0.050(4) \\
		 &  & 1273.66(15) & (5/2$^-$) & 2005.5(4) & 0.013(2) \\
	     &  & 310.44(5) & 3/2$^-$ & 2969.1(2) & 0.16(2) \\
		 &  & 63.46(6) & 1/2$^-$ & 3215.9(2) & 0.40(4) \\
		 &  & 0 & 5/2$^-$ & 3279.2(2) & 0.019(3) \\
3410.64(16)& (3/2$^+$,5/2$^+$) & 1920.41(8) & 5/2$^+$ & 1490.0(4) & 0.0098(15) \\
	     &  & 1417.71(7)& 1/2$^-$ & 1992.8(4) & 0.021(2) \\
		 &  & 692.36(5) & 3/2$^-$ & 2718.4(6) & 0.033(4) \\
		 &  & 310.44(5) & 3/2$^-$ & 3100.2(2) & 0.19(2) \\
3451.70(16)& (3/2$^+$,5/2$^+$) & 1920.41(8) & 5/2$^+$ & 1531.3(6) & 0.011(2) \\
		 &  & 1417.71(7)& 1/2$^-$ & 2033.7(5) & 0.010(2) \\
		 &  & 692.36(5) & 3/2$^-$ & 2759.3(2) & 0.094(9) \\
		 &  & 0 & 5/2$^-$ & 3451.7(4) & 0.024(3) \\
3509.18(14) & 3/2$^+$* & 1920.41(8) & 5/2$^+$ & 1588.7(2)& 0.30(2) \\
		 &  & 692.36(5) & 3/2$^-$ & 2817.8(6) & 0.111(10) \\
		 &  & 0 & 5/2$^-$ & 3509.0(2) & 0.062(7) \\ 
3962.63(19)& (3/2$^+$,5/2$^+$) & 2168.79(18) &  & 1793.8(5) & 0.026(2) \\
		 &  & 1417.71(7) & 1/2$^-$ & 2545.3(8) & 0.014(2) \\
		 &  & 1273.66(15) & (5/2$^-$) & 2688.9(3) & 0.040(4) \\
		 &  & 692.36(5) & 3/2$^-$ & 3269.8(5) & 0.022(3) \\
		 &  & 310.44(5) & 3/2$^-$ & 3652.0(5) & 0.083(9) \\
		 &  & 0 & 5/2$^-$ & 3962.8(5) & 0.19(2) \\
4001.49(18) $new$  & (1/2$^-$)* & 2711.27(8)& 3/2$^+$ & 1290.1(5) & 0.0184(14) \\
		 &  & 2168.79(18) &   & 1832.7(5) & 0.0076(10) \\
		 &  & 692.36(5) & 3/2$^-$ & 3309.3(3) & 0.23(2) \\
		 &  & 310.44(5) & 3/2$^-$ & 3691.0(5) & 0.35(4) \\
		 &  & 63.46(6) & 1/2$^-$ &  3937.6(5) & 0.039(5) \\
		 &  & 0 & 5/2$^-$ & 4000.8(5) & 0.014(2) \\
4344.65(15)& 1/2$^+$ & 1417.71(7) & 1/2$^-$ & 2927.1(5) & 0.099(8) \\
		 &  & 692.36(5) & 3/2$^-$ & 3652.1(2) & 0.106(10) \\
		 &  & 310.44(5) & 3/2$^-$ & 4033.7(5) & 0.054(6) \\
4391.81(18)& 3/2$^+$,5/2$^+$ & 2146.85(15)& 3/2$^-$ & 2244.8(5)& 0.0046(9) \\
		 &  & 1920.41(8) & 5/2$^+$ & 2471.6(8) & 0.0048(4) \\
		 &  & 1417.71(7) & 1/2$^-$ & 2974.5(8) & 0.025(2) \\
		 &  & 310.44(5) & 3/2$^-$ & 4080.8(5) & 0.024(3) \\
		 &  & 0 & 5/2$^-$ & 4391.6(5) & 0.117(13) \\		 
4508.0(4)&  & 2168.79(18) &  & 2338.9(5) & 0.025(2) \\
4544.49(22) &  & 2901.79(11)& 3/2$^+$,5/2$^+$ & 1642.9(5) & 0.026(2) \\
		 &  & 2324.20(6)& 3/2$^{(-)}$ & 2220(1) & 0.0147(13) \\
		 &  & 2146.85(15)& 3/2$^-$ & 2397.8(8) & 0.022(2) \\
		 &  & 1417.71(7) & 1/2$^-$ & 3126.7(6) & 0.017(2) \\
		 &  & 692.36(5) & 3/2$^-$ & 3851.8(5) & 0.031(4) \\		 
		 &  & 310.44(5) & 3/2$^-$ & 4233.5(8) & 0.063(7) \\	
		 &  & 0 & 5/2$^-$  & 4544.3(5) & 0.024(3) \\			 	 
4655.32(22)& 3/2$^+$,5/2$^+$ & 1920.41(8) & 5/2$^+$ & 2734.1(5)& 0.0079(14) \\
		 &  & 1417.71(7) & 1/2$^-$ & 3237.1(5) & 0.021(2) \\
		 &  & 1273.66(15) & (5/2$^-$) & 3381.9(5) & 0.0087(13) \\		 
		 &  & 310.44(5) & 3/2$^-$ & 4345.2(5) & 0.090(10) \\			 
		 &  & 0 & 5/2$^-$  & 4655.5(5) & 0.091(11)\\			 	 		 
		 
6098.28(11)& 1/2$^+$ & 4655.32(22)& 3/2$^+$,5/2$^+$ & 1442.9(8)& 0.23(2)\\
	   	 &  & 4544.49(22) & & 1553.6(5) & 0.199(10) \\
	   	 &  & 4508.0(4) & & 1589.9(5) & 0.031(2) \\
	   	 &  & 4391.81(18)& 3/2$^+$,5/2$^+$ & 1706.4(2) & 0.198(15) \\
	   	 &  & 4344.65(15)& 1/2$^+$  & 1753.5(2) & 0.250(14) \\
	   	 &  & 4001.49(18) & (1/2$^-$) & 2096.6(5) & 0.68(4) \\
	   	 &  & 3962.63(19) & (3/2$^+$,5/2$^+$) & 2135.6(6) & 0.41(2) \\
	   	 &  & 3509.18(14) & 3/2$^+$ & 2589.0(5) & 0.46(3) \\ 
	   	 &  & 3451.70(16)& (3/2$^+$,5/2$^+$) & 2646.6(5) & 0.164(11) \\
	   	 &  & 3410.64(16) & (3/2$^+$,5/2$^+$) & 2687.7(5) & 0.24(2) \\
	   	 &  & 3279.40(10)& 3/2$^+$,5/2$^+$  & 2819.2(5) & 0.70(5) \\
	   	 &  & 3105.68(24) & (1/2$^-$,3/2$^-$) & 2992.7(8) & 0.036(4) \\
	   	 &  & 3044.27(20) &  & 3054.4(8) & 0.010(2) \\
	   	 &  & 3014.43(14) &  & 3084.0(8) & 0.108(8) \\
	   	 &  & 2901.79(11) & 3/2$^+$,5/2$^+$ & 3196.7(5) & 0.126(9) \\
	   	 &  & 2793.05(10) & 5/2$^+$ & 3305.2(5) & 0.099(8) \\
	   	 &  & 2711.27(8)& 3/2$^+$ & 3387.0(5) & 3.2(2) \\	   	 
	   	 &  & 2324.20(6)& 3/2$^{(-)}$  & 3773.9(5) & 1.67(8) \\	   	 
	   	 &  & 2146.85(15)& 3/2$^-$  & 3951.5(5) & 5.6(4) \\	   	 
	   	 &  & 1920.41(8) & 5/2$^+$ & 4177.8(8) & 0.70(5) \\	   	 
	     &  & 1417.71(7) & 1/2$^-$ & 4680.3(5) & 17(2) \\	   	  
	  	 &  & 692.36(5) & 3/2$^-$ & 5405.6(3) & 46(6) \\	   	  
	  	 &  & 310.44(5) & 3/2$^-$ & 5787.7(5) & 86(12) \\
	  	 &  & 63.46(6) & 1/2$^-$  & 6034.8(5) &  \\ 	   	  
	   	 
\hline
\hline
\end{longtable*}
\renewcommand\baselinestretch{1.15}\selectfont

\subsection{\label{sec:angcorr}Angular correlations, multipolarity determinations, and spin-parity assignments}

Multipolarity mixing ratios of $\gamma$ transitions and spin and parity assignments for a number of excited states in $^{65}$Ni were established by employing a  $\gamma\gamma$ angular correlation analysis.

Given two $\gamma$ rays emitted in the same cascade at an angle $\theta$ between each other, the angular correlation $W(\theta)$ between them is generally expressed as a superposition of Legendre polynomials $P_n$:
\begin{equation}
\label{eq:AngCor}
W(\theta)=  A_0 [ 1+A_2 P_2(\cos\theta)+A_4 P_4(\cos\theta)]
\end{equation}
where $A_0$ is a normalization coefficient and $A_n$ ($n=2,\,4$) are the product of two coefficients, $A_n(1)$ and $A_n(2)$, depending on the character of the transitions and the spin of the levels involved~\cite{Morinaga}:
\begin{equation}
\label{eq:AngCorCoeff}
A_n = q_n A_n(1) A_n(2),\quad \mathrm{with}\,n = 2,4.
\end{equation}
Here $q_n$ indicates attenuation terms, which take into account the finite solid angle of the detectors.

The number of angle pairs of the FIPPS array considered in the present study is 19. The attenuation coefficients $q_n$ for the apparatus were obtained considering different pairs of $\gamma$ rays from a $^{152}$Eu source, with known character and multipolarity mixing ratios. The established values are $q_2$ = 0.80(10) and $q_4$ = 0.60(10). 

The multipolarity mixing ratio $\delta_{\gamma}$ for a $\gamma$ transition is defined as follows~\cite{Morinaga}:
\begin{equation}
    \delta = \frac{\langle j_f || \lambda' || j_i \rangle}{\langle j_f || \lambda || j_i \rangle}
\end{equation}
where $\lambda$ is the lowest possible multipolarity of the photon $\gamma$ emitted in the transition between the states with initial spin $j_i$ and final spin $j_f$, and $\lambda'=\lambda+1$.
The mixing ratio value of a certain transition was extracted from the fit of  angular correlation experimental data when the mixing ratio of the other transition of the considered pair was known, so that either $A_n(1)$ or $A_n(2)$ in eq.~\ref{eq:AngCorCoeff} is fixed, for each $n$. A fit of the experimental data was thus performed giving the experimental $A_2^\text{exp}$ and $A_4^\text{exp}$ values, which could then be compared to the corresponding theoretical analytical expressions (in the following indicated as $A_2(\delta)$ and $A_4(\delta)$), as a function of $\delta$. The $\delta$ value was obtained minimizing the $\chi^2$ function defined as:
\begin{equation}
    \chi^2= \bigg(\frac{A_2^\text{exp}-A_2(\delta)}{\Delta A_2^\text{exp}}\bigg)^2+\bigg(\frac{A_4^\text{exp}-A_4(\delta)}{\Delta A_4^\text{exp}}\bigg)^2
\end{equation}
where $\Delta A_2^{\text{exp}}$ and $\Delta A_4^{\text{exp}}$ are the uncertainty of $A_2^{\text{exp}}$ and $A_4^{\text{exp}}$.

We started the analysis by establishing the mixing ratios of the two most intense transitions in the decay scheme, namely the 310- ($3/2^-\rightarrow 5/2^-$) and 629-keV ($3/2^-\rightarrow 1/2^-$) $\gamma$ rays. These two transitions were subsequently used to determine the mixing ratios of other lines considered in pairs with them. 
A few pairs of non-consecutive transitions were considered. In these cases, to estimate the uncertainty introduced by a possible de-orientation caused by the intermediate transitions, the approach introduced in Ref.~\cite{Cieplicka_PRC94_2016} was applied. A de-orientation was simulated by increasing the angular correlation attenuation in eq.~\ref{eq:AngCorCoeff}, up to 50\% for each cascade, and the corresponding variation of the resulting mixing ratio was adopted as a systematic uncertainty.  When multiple pairs could be considered for the investigated transition (i.e., for the primary 3774- and 3952-keV $\gamma$~rays), $\delta$ was taken as an average of the values resulting from the fit of individual pairs. In this case, the uncertainty was calculated taking into account the variation in $\delta$ values obtained from the analysis of the different pairs.
 
The information on angular correlation coefficients $A_2^\text{exp}, A_4^\text{exp}$ and mixing ratios $\delta$ is summarized in Tab.~\ref{tab:AngCor} for all the considered $\gamma$ cascades. In the following paragraphs, we first discuss the angular correlation analysis performed to establish the multipolarity mixing ratios for the  310- and 629-keV transitions. Further, the spin assignments for a number of states are deduced. 

\renewcommand\baselinestretch{1.2}\selectfont  
\begin{table*}
\caption{Experimental $A_2^\text{exp}$ and $A_4^\text{exp}$ coefficients and $\delta$ mixing ratios for each pair of $\gamma$ rays $E_{\gamma1}$-$E_{\gamma2}$ studied in this work, together with corresponding analytical $A_2$ and $A_4$ values. The spins of the levels involved are listed in the second column. In the case of non-consecutive $\gamma$-ray cascades, all the four spins of the levels involved are given (see text for details). For the mixing ratio values, the sign convention defined by Krane and Steffen has been used~\cite{KraneSteffen1970}.}

\small
\centerfloat
\setlength{\tabcolsep}{0.1em}
\begin{tabular}{lccSSSSSS}
\hline
\hline
$\boldsymbol{E_{\gamma1}}$-$\boldsymbol{E_{\gamma2}}$ & $\boldsymbol{J^\pi_i \rightarrow J^\pi_m \rightarrow J^\pi_f}$ & \textbf{Multipolarity} & {$\boldsymbol{A_2^\text{exp}}$} & {$\boldsymbol{A_4^\text{exp}}$} & {$\boldsymbol{A_2}$} & {$\boldsymbol{A_4}$} & {$\boldsymbol{\delta_1}$} & {$\boldsymbol{\delta_2}$} \\
\hline
1107-310 & $1/2^-\rightarrow3/2^-\rightarrow5/2^-$ & $M1$(+$E2$)-$M1$(+$E2$) & 0.176(5) & 0.001(5) & 0.17 & 0.00 & 0.017(8) & 0.191(13) \\
1228-629 & $5/2^+\rightarrow3/2^-\rightarrow1/2^-$ & $E1$(+$M2$)-$M1$(+$E2$) & 0.09(2) & -0.02(3) & 0.08 & 0.00 & -0.09(4)  & 0.052(11) \\
2097-310\footnote{\label{2}non-consecutive $\gamma$ rays} & $1/2^+\rightarrow(1/2^-),\,$ & $E1$-$M1$(+$E2$) & 0.02(2) & 0.00(2) & 0.00 & 0.00 & 0.00(5) & 0.191(13) \\
 & $3/2^-\rightarrow5/2^-$ & & & & & & &\\
2589-629\footref{2} & $1/2^+\rightarrow3/2^+,\,$ & $M1$(+$E2$)-$M1$(+$E2$) & 0.15(2) & 0.01(4) & 0.14 & 0.00 & -0.1(2)\,or & 0.052(11) \\
 & $3/2^-\rightarrow1/2^-$ &  &  &  & &  & 2.1(7) & \\
3774-310\footref{2} & $1/2^+\rightarrow3/2^{(-)},\,$ & $E1$(+$M2$)-$M1$(+$E2$) & 0.08(2) & 0.03(2) & 0.10 & 0.00 & -0.1(2)\footnote{\label{5}value determined averaging results from the transition pairs 3774-310 and 3774-629~keV} & 0.191(13) \\
& $ 3/2^-\rightarrow5/2^-$ & & & & & & &\\
3774-629\footref{2} & $1/2^+\rightarrow3/2^{(-)},\,$ & $E1$(+$M2$)-$M1$(+$E2$) & 0.15(2) & 0.01(3) & 0.12 & 0.00 & -0.1(2)\footref{5} & 0.052(11)\\
& $3/2^-\rightarrow1/2^-$ & & & & & & & \\
3952-310\footref{2} & $1/2^+\rightarrow3/2^-,\,$ & $E1$(+$M2$)-$M1$(+$E2$) & 0.06(2) & 0.05(3) & 0.08 & 0.00 & -0.1(2)\footnote{\label{6}value determined averaging results from the transition pairs 3952-310 and 3952-629~keV} & 0.191(13) \\
& $3/2^-\rightarrow5/2^-$ & & & & & & &\\
3952-629\footref{2} & $1/2^+\rightarrow3/2^-,\,$ & $E1$(+$M2$)-$M1$(+$E2$) & 0.13(2) & 0.04(3) & 0.11 & 0.00 & -0.1(2)\footref{6} & 0.052(11) \\
 & $3/2^-\rightarrow1/2^-$ & & & & & & & \\
5406-629\footnote{\label{4}cascade used to determine the $\delta_{629}$ mixing ratio of the 629-keV transition} & $1/2^+\rightarrow3/2^-\rightarrow1/2^-$ & $E1$-$M1$(+$E2$) & 0.213(10) & -0.02(2) & 0.20 & 0.00 &  0.00(5)\footnote{\label{1}value assumed on the basis of angular momentum selection rules (see text for details)} & 0.052(11)\\
5406-692 & $1/2^+\rightarrow3/2^-\rightarrow5/2^-$ & $E1$-$M1$(+$E2$) & 0.066(10) & 0.005(14) & 0.07 & 0.00 &  0.00(5)\footref{1} & 0.03(2) \\
5788-310\footnote{\label{3}cascade used to determine the $\delta_{310}$ mixing ratio of the 310-keV transition} & $1/2^+\rightarrow3/2^-\rightarrow5/2^-$ & $E1$-$M1$(+$E2$) & 0.171(2) & -0.004(4) & 0.16 & 0.00 &  0.00(5)\footref{1} & 0.191(13)\\
\hline
\hline
\end{tabular}
\label{tab:AngCor}
\end{table*}

\renewcommand\baselinestretch{1.15}\selectfont

\begin{figure}
\centering
\includegraphics[width=\linewidth]{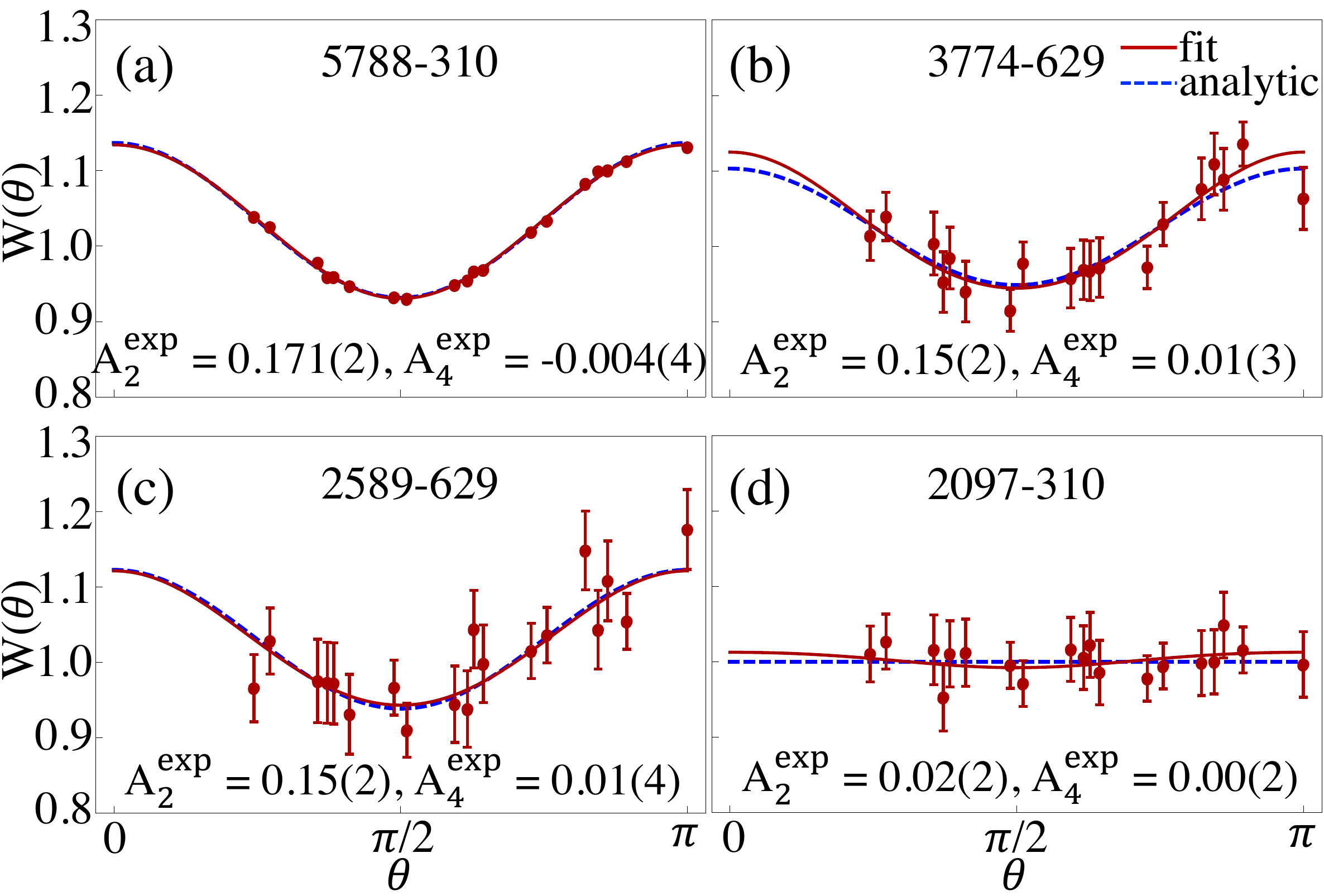}
\caption{Examples of angular correlations for pairs of $\gamma$ rays in $^{65}$Ni. The red solid line is the function $W(\theta)$ (see Eq.~\ref{eq:AngCor}) fitted to the experimental points, with $A_2^\text{exp}$ and $A_4^\text{exp}$ values  given in each panel. The blue dashed line is the curve calculated from the analytical expressions $A_2(\delta)$ and $A_4(\delta)$ by taking the $\delta$ mixing ratio which resulted from the analysis discussed above. Such curve reflects the quality of the mixing ratio fit, provided the spins of the states involved are fixed. Detector combinations at the smallest angles are not considered to avoid systematic errors caused by enhancement of the Compton background.}
\label{fig:AngCorPlot} 
\end{figure}

\subsubsection{Multipolarity mixing ratio of the 310-keV and 629-keV transitions}
\label{subsec:mix310_629_692}

\paragraph*{\textbf{310~keV, $\mathbf{3/2^-\rightarrow 5/2^-}$.}}
The 310-keV ground state transition between $3/2^-$ and $5/2^-$ states is expected to have an $M1$+$E2$ character. 
In order to determine its multipolarity mixing ratio $\delta_{310}$, the $\gamma$-ray cascade 5788-310~keV was considered. The angular correlation plot for this cascade is presented in panel~(a) of Fig.~\ref{fig:AngCorPlot}. 
For the  5788-keV ($1/2^+\rightarrow 3/2^-$) primary $\gamma$-ray transition, a dominant $E1$ character was assumed on the basis of angular momentum selection rules, with a maximum $M2$ mixing contribution of $\delta_{5788}<0.05$, which is consistent with the work of Ref.~\cite{Cieplicka_PRC93_2016}. The $\delta_{310}$ value obtained with such assumption (i.e., averaging the solutions considering the lower and upper limits $\delta_{5788}=0.00$ and $\delta_{5788}=0.05$) is 0.191(13).

\paragraph*{\textbf{629~keV, $\mathbf{3/2^-\rightarrow 1/2^-}$.}}
For the 629-keV transition, depopulating the 692-keV state, an $M1$+$E2$ character was expected. 
To determine the multipolarity mixing ratio $\delta_{629}$, the cascade 5406-629~keV was considered. 
Similarly to the previous case, for the 5406-keV primary transition a dominant $E1$ character was assumed, with a maximum $M2$ mixing contribution of $\delta_{5406}<0.05$. The two mixing ratio solutions for the 629-keV transition are $\delta_{629}=0.052(11)$ and $\delta_{629}=-1.95(6)$. To constrain this result, the $\approx$~0.7~ps lifetime of the 692-keV level, provided by a 1-neutron transfer reaction $^{64}$Ni($^{13}$C,$^{12}$C)$^{65}$Ni study performed at sub-barrier Coulomb energy at IFIN-HH (Bucharest, Romania)~\cite{Sferrazza}, was used. For the mixing ratio $\delta_{629}=-1.95(6)$, the resulting reduced transition probability value would be $B(E2)\approx480$~W.u., which is unrealistically large. On the other hand, $\delta_{629}=0.052(11)$ gives $B(E2)\approx1$~W.u., which is a reasonable value.
Having this result, we analyzed the angular correlation between the 629-keV and 1228-keV $\gamma$ rays, which yielded a multipolarity mixing ratio for the 1228-keV transition $\delta_{1228}=-0.09(4)$, pointing to its $E1$+$M2$ character.

\subsubsection{2324-keV level: $J^{\pi}=3/2^{(-)}$ spin assignment}
In order to assign spin and parity to the 2324-keV level, never determined before, the angular correlation between the primary 3774-keV $\gamma$ ray feeding this level and the 310- and 629-keV lower-lying transitions were studied. 
The analysis pointed to a positive $A_2^\text{exp}$~coefficient for both pairs, as shown, for example, for the 3774-629~keV case in panel~(b) of Fig.~\ref{fig:AngCorPlot}. 
Given the spin of the capture state $J_{6098}=1/2$, the most probable spin assignment for the 2324-keV level is 3/2, because $J_{2324}=1/2$ would imply an isotropic angular correlation plot, while $J_{2324}=5/2$ would give negative $A_2$~coefficients.
Subsequently, the angular correlations for both pairs were used to establish the multipolarity mixing ratio of the 3774-keV transition, for which a dipole+quadrupole character was expected. A solution common to both cascades is $\delta_{3774}=-0.1(2)$. 
We tentatively propose a negative parity for the 2324-keV level, since the 3774-keV primary $\gamma$ transition is more likely to have an $E1$+$M2$ character ($1/2^+\rightarrow3/2^-$) with low mixing, rather than $M1$+$E2$ ($1/2^+\rightarrow3/2^+$), similarly to the other primary transitions reported in Tab.~\ref{tab:AngCor}.

\subsubsection{3509-keV level: $J^{\pi}=3/2^+$ spin assignment}
In earlier works, the 3509-keV level was reported as $3/2^+$, $5/2^+$~\cite{ENSDFurl}. To restrict this assignment, the non-consecutive $\gamma$-ray cascade 2589-629~keV was analyzed, where the 2589-keV transition is a primary $\gamma$ ray populating the 3509-keV level (see panel~(c) of Fig.~\ref{fig:AngCorPlot}).
The other non-consecutive cascade 2589-310~keV could not be used, since the 2589-keV transition is a doublet, with both the transitions of the doublet in coincidence with the 310-keV line. 
Only the $J^{\pi}_{3509}=3/2^+$ assignment is possible to obtain fit convergence.
This spin-parity assignment implies an $M1$+$E2$ character for the 2589-keV transition. By fitting the angular correlation data points, the two solutions $\delta_{2589}=2.1(7)$ and $-0.1(2)$ are found for the mixing ratio of the 2589-keV transition. No other experimental information is available to further constrain the $\delta_{2589}$ value. 

\subsubsection{4001-keV level: $J^{\pi}=(1/2^-)$ spin assignment}
A tentative spin assignment for the newly observed 4001-keV level is proposed, through the study of the angular correlation of the non-consecutive $\gamma$-ray cascade 2097-310~keV, where 2097~keV is the primary $\gamma$~ray. 
The spins considered as possible options are $J_{4001}=1/2$, 3/2, 5/2. The angular correlation appears quite isotropic (see panel (d) of Fig.~\ref{fig:AngCorPlot}), which is not compatible with $J_{4001}=5/2$. 
In turn, the spin-parity assignment of $1/2^-$, implying a dominant $E1$ character for the 2097-keV primary  $\gamma$~ray (with a maximum $M2$ mixing of $\delta_{2097}<0.05$, similarly to the cases discussed above), is consistent with the angular correlation data and the direct decay branch to the 5/2$^-$ ground state. The spin $J_{4001}=3/2$ cannot be firmly excluded, although it would require a large mixing for the primary 2097-keV $\gamma$~ray. We adopt the tentative spin-parity assignment of ($1/2^-$) for the 4001-keV level.

\subsubsection{Other angular correlations considered}
Other cascades of intense $\gamma$~rays were considered in the angular correlation analysis. In particular, on the basis of spin-parity assignments from earlier works~\cite{ENSDFurl}, the $1/2^+\rightarrow3/2^-$ 3952-keV transition is expected to have $E1$(+$M2$) character. In our analysis, by considering the cascades 3952-310 and 3952-629~keV, the mixing ratio $\delta_{3952}=-0.1(2)$ was obtained, similarly to other primary $\gamma$ rays considered, which have a dominant $E1$ character. 

Furthermore, the cascade 5406-692~keV was analyzed. As discussed above, the primary 5406-keV transition is a pure $E1$ (with a maximum $M2$ mixing contribution of $\delta_{5406}<0.05$), while the 692-keV ground state transition depopulates the 692-keV state with a $\approx$~0.7~ps lifetime~\cite{Sferrazza}. 
The solutions for the $M1$/$E2$ mixing ratio of the $3/2^-\rightarrow5/2^-$ 692-keV transition are $\delta_{692}=0.03(2)$ and $-5.3$(4), which lead to B(E2) reduced transition probabilities of $\approx$~0.06~W.u.~and $\approx$~67~W.u., respectively. Assuming a low collectivity for the low-lying states of $^{65}$Ni, as predicted by shell-model calculations (later discussed), the first solution $\delta_{692}=0.03(2)$ is adopted.

The same kind of argument was used to define the multipolarity mixing ratio for the $1/2^-\rightarrow3/2^-$ ($M1$+$E2$) 1107-keV transition, depopulating the 1/2$^-$ state at 1418~keV, with lifetime upper limit of 300~fs, as determined in the complementary experiment of Ref.~\cite{Sferrazza}. The angular correlation study of the 1107-310~keV pair gives the mixing ratio values $\delta_{1107}=1.66(2)$ and $0.017(8)$, which correspond to $B(E2)>72$~W.u.~and $B(E2)>0.03$~W.u., respectively. This points to $\delta_{1107}=0.017(8)$ for the mixing ratio of the 1107-keV transition. 

\section{\label{sec:theor}Comparison with Monte Carlo shell-model calculations}
\begin{figure*}
\centering
\includegraphics[width=0.85\textwidth]{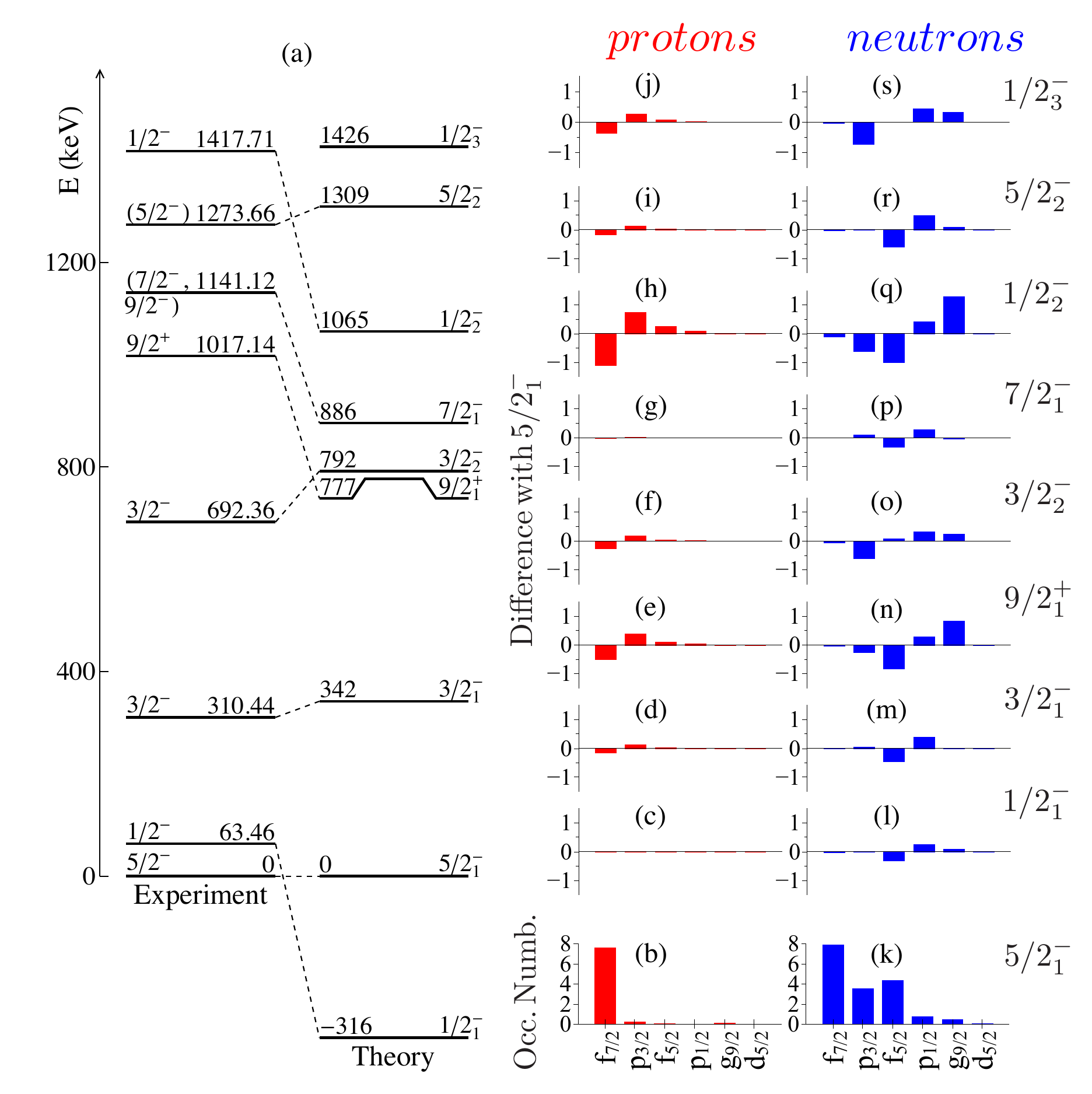}
\caption{(a) Excitation energy spectrum of $^{65}$Ni, up to 1417~keV, from experiment and MCSM calculations. (b) MCSM predictions for the proton occupation number of the 5/2$^-_1$ state. (c)-(j) Calculated differences with respect to the 5/2$^-_1$ state for proton occupancy, for each excited state. (k) MCSM predictions for the neutron occupation number of the 5/2$^-_1$ state. (l)-(s) Calculated differences with respect to the 5/2$^-_1$ state for neutron occupancy, for each excited state. The excited states are indicated at the right side of the panels.}
\label{fig:occNumb} 
\end{figure*}

The experimental level scheme of $^{65}$Ni has been compared, up to 1.4~MeV, with Monte Carlo Shell Model (MCSM) results~\cite{Tsunoda}. The calculations are performed considering an inert core of $^{40}$Ca and the proton and neutron single-particle orbits $f_{7/2}$, $p_{3/2}$, $f_{5/2}$, $p_{1/2}$, $g_{9/2}$ and $d_{5/2}$ (i.e., the $pf$-$g_{9/2}$-$d_{5/2}$ model space).

Fig.~\ref{fig:occNumb}(a) shows the comparison between experimental results, in terms of energy and spin/parity of each state, and predictions from the MCSM. All excited states in the energy region here considered have negative parity, apart from the $J^\pi$=9/2$^+$ state, located at an experimental excitation energy of 1017.1~keV. It is found that each experimental state has a theoretical counterpart, although with a discrepancy in energy up to 400~keV.  An inversion between the ground state and the first excited state is also observed.

\begin{figure}
\centering
\includegraphics[width=\linewidth]{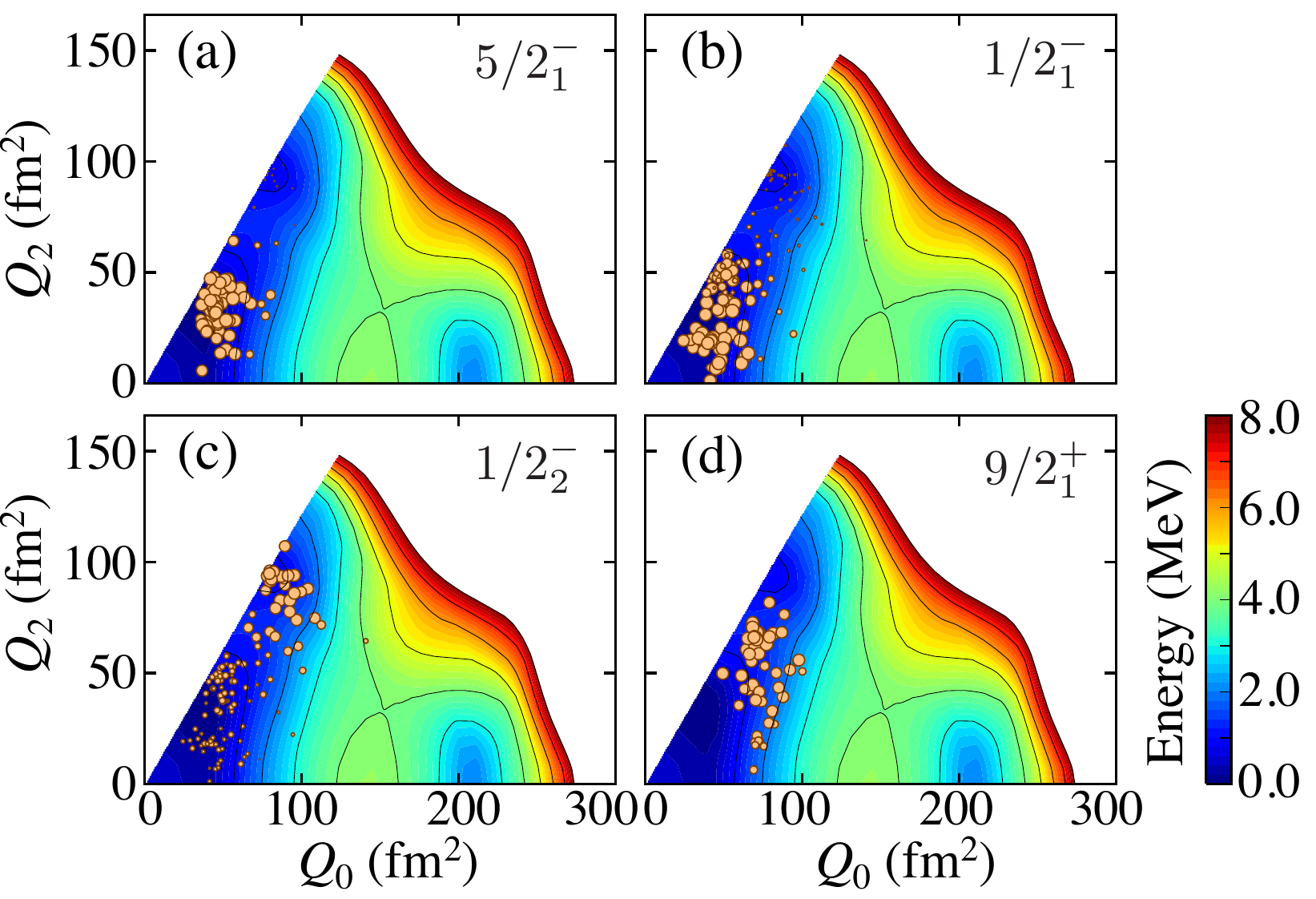}
\caption{(Color online) Potential energy surfaces of $^{65}$Ni, obtained via constrained Hartree-Fock calculations. Circles on the energy surface represent shapes of MCSM basis vectors according to a T-plot analysis for the states 5/2$^-_1$, 1/2$^-_1$, 1/2$^-_2$ and 9/2$^+_1$, shown in panels (a), (b), (c) and (d), respectively (see text for further details).}
\label{fig:Tplot} 
\end{figure}
Information on the configurations of the calculated states can be obtained by inspecting the occupation numbers predicted by theory for the proton and neutron single particle orbitals, as shown in Figs.~\ref{fig:occNumb}(b) to (s). Figures~\ref{fig:occNumb}(b) and (k) give the occupation numbers for the 5/2$^-_1$ state, which is the experimental ground state. In this case, while the protons are mainly in the $f_{7/2}$ orbital, the neutrons occupy the $f_{7/2}$, $p_{3/2}$ and $f_{5/2}$ single-particle states, with smaller contributions from $p_{1/2}$, $g_{9/2}$ and $d_{5/2}$. The associated intrinsic shape is predicted spherical. Figures~\ref{fig:occNumb}(c)-(j) and (l)-(s) of  show the differences, in proton and neutron occupation numbers, between each excited state and the 5/2$^-_1$ ground state. All states, apart from 1/2$^-_2$ and 9/2$^+_1$, are characterized by small differences ($<$1 unit of occupation number), thus indicating similar wave function compositions as the spherical 5/2$^-_1$ ground state. 
On the contrary, in the case of the 1/2$^-_2$ state, a sizable enhancement in the occupation of the neutron $g_{9/2}$ and proton $p_{3/2}$ and $f_{5/2}$ orbitals, with respect to 5/2$^-_1$, is found.
For the 9/2$^+_1$ state, naturally the $g_{9/2}$ neutron orbital is occupied. In the neutron-rich Ni isotopes, such occupation pattern is characteristic of deformed structures~\cite{Tsunoda}. 
A visualization of the deformation associated with a given MCSM state is offered by the so called T-plot representation~\cite{Otsuka, Tsunoda}, as shown in Fig.~\ref{fig:Tplot} for the 5/2$^-_1$, 1/2$^-_1$, 1/2$^-_2$ and 9/2$^+_1$ states of $^{65}$Ni. Here, circles on the potential energy surface indicate the projection of the MCSM eigenstates on the quadrupole moment coordinates.
It is clearly seen that for the 5/2$^-_1$ ground state circles are concentrated around $Q_0 = Q_2 = 0$, indicating a spherical shape. The same feature is shown by the 1/2$^-_1$ state (Fig.~\ref{fig:Tplot}(b)), pointing to a spherical shape, as already suggested by the similarity of occupation numbers with the 5/2$^-_1$ experimental ground state discussed above.
In turn, circles on the T-plot corresponding to the 1/2$^-_2$ and 9/2$^+_1$ states are located around the oblate minimum. 

\section{\label{sec:conclus}Conclusions}

The low-spin structure of the $^{65}$Ni nucleus was investigated in the neutron capture experiment $^{64}$Ni(n,$\gamma$)$^{65}$Ni performed at ILL, Grenoble, with the FIPPS array. 24~primary $\gamma$-rays were observed (15~new) and a total of 87~new $\gamma$-ray transitions were found, thus increasing considerably our knowledge on the low-spin structure of this nucleus. In total 28~discrete states (2~new) were located below the 6098.3~keV neutron binding energy.

The lower excitation energy part of the level scheme, up to about 1.4~MeV, was compared to MCSM calculations, performed in the $pf$-$g_{9/2}$-$d_{5/2}$ neutron and proton model space. The calculations indicate a spherical character of the considered states, apart from the second 1/2$^-_2$ excitation, at 1417.8~keV, and the first 9/2$^+_1$ excitation, at 1017.1~keV. In these cases, a significant occupation of the proton $p_{3/2}$ and $f_{5/2}$ and neutron $g_{9/2}$ orbitals is found, which is typical of deformed structure in the neutron-rich Ni region. A more detailed analysis of the state deformation, offered by the T-plot, points to an oblate shape for both the 1/2$^-_2$ and 9/2$^+_1$ states. According to MCSM calculations, prolate deformed structures are expected to occur at higher excitation energies (i.e.~around 3~MeV), where the comparison between theory and experiment is more complex, due to the experimental uncertainty in spin-parity assignments.

\section{Acknowledgements}
This work was supported by the Italian Istituto
Nazionale di Fisica Nucleare, by the Polish National Science Centre under Contract No.~UMO-2015/14/m/ST2/00738 (COPIN-INFN Collaboration), by the Fonds de la Recherche Scientifique - FNRS under Grant Number 4.45.10.08 and by the Romanian Nucleu Program PN 19 06 01 02. The MCSM calculations were performed on K computer at RIKEN AICS (hp160211,
hp170230, hp180179, hp190160). This work was supported in part by MEXT as "Priority Issue on Post-K computer" (Elucidation of the Fundamental Laws and Evolution of the Universe) and JICFuS.

\bibliography{biblio}

\end{document}